\documentclass[aps]{revtex4}
\usepackage{epsfig}
\usepackage{graphicx}
\begin{document}
\title{Relativistic Equation of state with short range
correlations}
\author{P.K. Panda}
\author{D.P. Menezes}
\affiliation{Depto de F\'isica-CFM, Universidade Federal de Santa
Catarina, CP. 476, 88040-900 Florian\'opolis-SC, Brazil}
\author{C. Provid\^encia}
\author{J. da Provid\^encia}
\affiliation{Centro de F\'isica Te\'orica - Dep. de F\'isica,
Universidade de Coimbra, P-3004 - 516 Coimbra, Portugal}
\begin{abstract}
Short range correlations are introduced using unitary correlation method
in a relativistic approach to the equation of state of the
infinite nuclear matter in the framework of the Hartree-Fock
approximation. The effect of the correlations in the
ground state properties of the nuclear matter is studied.
\end{abstract}
\maketitle
In this paper we introduce short range correlations 
in a relativistic approach to the description of nuclear matter for
the first time. Although there are several procedures which
may be used to introduce short range correlations into the model wave
function, we work with the unitary operator method as
proposed by Villars \cite {villars}. There are several advantages in using an
unitary model operator. In particular, one automatically guarantees that the
correlated state is normalized.
The general idea of introducing short range correlations in systems
with short range interactions exists for a long
time \cite{jp,feldmeier} but has not been pursued for the
relativistic case. 

Nonrelativistic calculations based on realistic NN potentials predict
equilibrium points that are not able to describe simultaneously the correct
binding energy and saturation density;
either the saturation
density is correct but the  binding energy is too small, or the correct binding
energy is obtained at a too high density \cite{Coester}.
In order to solve this problem a repulsive potential or 
density-dependent repulsive mechanism \cite{3-body} is included.
Due to Lorentz covariance and self-consistency, relativistic mean
field theories \cite{walecka} include automatically contributions which are
equivalent to $n$-body repulsive potentials in non-relativistic approaches.

In non-relativistic
models the interaction arises from the interplay between a long
range attraction and a very strong short range repulsion and it is 
indispensable to take short range correlations into account.
In relativistic mean field models, 
the parameters are phenomenological, fitted to the
saturation properties of nuclear matter.  Short
range correlation effects are may be included to some extent in the model 
parameters. However, we want to study the  consequences  of 
taking these effects into account explicitly \cite{corr}.

We start with the effective Hamiltonian as
\begin{equation}
H=\int \psi^\dagger_\alpha (\vec x)(-i\vec\alpha\cdot\vec\nabla
+\beta M)_{\alpha\beta}\psi_\beta(\vec x)~d\vec x
+\frac{1}{2}\int\psi_\alpha^\dagger (\vec x)\psi_\gamma^\dagger (\vec y)
V_{\alpha\beta,\gamma\delta}(|\vec x -\vec y|)
\psi_\delta(\vec y)\psi_\beta(\vec x)~d\vec x~d\vec y
\end{equation}
with
\begin{equation}
V_{\alpha\beta,\gamma\delta}(r)=
(\beta)_{\alpha\beta}(\beta)_{\gamma\delta}V_\sigma(r)
+\left(\delta_{\alpha\beta}\delta_{\gamma\delta}
-\vec\alpha_{\alpha\beta}\cdot\vec\alpha_{\gamma\delta}\right)V_{\omega}(r)
\end{equation}
where
\begin{equation}
V_\sigma(r)=-\frac{g_\sigma^2}{4 \pi}\frac{e^{-m_\sigma
r}}{r},\quad\quad V_{\omega}(r)=\frac{g_\omega^2}{4
\pi}\frac{e^{-m_\omega r}}{r},
\end{equation}
and $\vec\alpha$ are the  Dirac-matrices.
In the above, $\psi$ is the nucleon field interacting through the scalar
and vector potentials.
The equal time quantization condition for the nucleons is given by
\begin{equation}
[\psi _{\alpha}(\vec x,t),\psi _\beta (\vec y,t)^{\dagger}]_{+}
=\delta _{\alpha \beta}\delta(\vec x -\vec y),
\end{equation}
where $\alpha$ and $\beta$ refer to the spin indices. The
field expansion for the nucleons
$\psi$ at time t=0  given by \cite{mishra}
\begin{equation}
\psi(\vec x)=\frac {1}{\sqrt{V}}\sum_{r,k} \left[U_r(\vec k)c_{r,\vec k}
+V_r(-\vec k)\tilde c_{r,-\vec k}^\dagger\right] e^{i\vec k\cdot \vec x} ,
\end{equation}
where $U_r$ and $V_r$ are
\begin{equation}
U_r(\vec k)=\left( \begin{array}{c}\cos\frac{\chi(\vec k)}{2}
\\ \vec \sigma \cdot\hat k\sin\frac{\chi(\vec k)}{2}
\end{array}\right)u_r~ ;~~ V_r(-\vec k)=\left(
\begin{array}{c}-\vec \sigma \cdot\hat k\sin\frac{\chi(\vec k)}{2}
\\ \cos\frac{\chi(\vec k)}{2}\end{array}\right)v_r~.
\end{equation}
For free spinor fields, we have $\cos\chi(\vec k)=M/\epsilon(\vec
k)$, $\sin\chi(\vec k)=|\vec k|/\epsilon(\vec k)$ with
$\epsilon(\vec k)=\sqrt{\vec k^2 + M^2}$. However, we will deal
with interacting fields so that we take the ansatz $\cos\chi(\vec
k)=M^*(\vec k)/\epsilon^*(\vec k)$, $\sin\chi(\vec k)=|\vec
k^*|/\epsilon^*(\vec k)$, with $\epsilon^*(\vec k)=\sqrt{\vec
{k^*}^2 + {M^*}^2(\vec k)}$, where $\vec k^*$ and $M^*(\vec k)$
are the effective momentum and effective mass, respectively. The
equal time anti-commutation conditions are 
\begin{equation}
[c_{r,\vec
k},c_{s,\vec k'}^\dagger]_{+}~=~ \delta _{rs}\delta_{\vec k,\vec
k'}~=~ [\tilde c_{r,\vec k},\tilde c_{s,\vec k'}^\dagger]_{+}~. 
\end{equation}
The vacuum $\mid 0\rangle$ is defined through $c_{r,\vec k}\mid
0\rangle=\tilde c_{r,\vec k}^\dagger\mid 0 \rangle=0$;
one-particle states are written $|\vec k,r\rangle=c_{r,\vec
k}^\dagger\mid 0\rangle$; two-particle and three-particle
uncorrelated states are written, respectively as $|\vec k,r;\vec
k',r'\rangle= c_{r,\vec k}^\dagger ~c_{r',\vec k'}^\dagger\mid
0\rangle$, and $|\vec k,r;\vec k',r'; \vec k'',r''\rangle=
c_{r,\vec k}^\dagger ~c_{r',\vec k'}^\dagger ~c_{r'',\vec
k''}^\dagger \mid 0\rangle,$ and so on.

We now introduce the short range correlation
through an unitary operator method. The correlated wave function\cite{jp1} is
$|\Psi\rangle=e^{i\Omega}|\Phi\rangle$ where $|\Phi\rangle$
is a Slater determinant and $\Omega$ is, in general, a $n$-body
Hermitian operator, splitting into a 2-body part, a 3-body part,
etc.. The expectation value of $H$ is
\begin{equation}
E=\frac{\langle \Psi| H|\Psi\rangle}{\langle \Psi |\Psi\rangle}=
\frac{\langle \Phi| e^{-i\Omega}~H~e^{i\Omega}|\Phi\rangle}{\langle \Phi|
\Phi\rangle}.
\end{equation}
In the present calculation, we only take into account two-body correlations. 
Let us denote the  two-body  correlated wave function by
\begin{equation}
|\overline{\vec
k,r;\vec k',r'}\rangle=e^{i\Omega}|{\vec k,r;\vec k',r'}\rangle\approx
f_{12}|{\vec k,r;\vec k',r'}\rangle
\end{equation}
where $f_{12}$ is the short range correlation factor, the
so-called Jastrow factor \cite{jastrow}. For simplicity, we consider
$f_{12}=f(\vec r_{12})$, $\vec r_{12}=\vec r_1-\vec r_2$,
and $f(r)=1-(\alpha +\beta r)~e^{-\gamma r}$
where $\alpha$, $\beta$ and $\gamma$ are parameters.
The important effect of the short range
correlations is the replacement, in the expression for the
ground-state energy, of the interaction matrix element
$\langle{\vec k,r;\vec k',r'}|V_{12}|{\vec k,r;\vec k',r'}\rangle$
by $\langle\overline{\vec k,r;\vec
k',r'}|V_{12}+t_1+t_2|\overline{\vec k,r;\vec
k',r'}\rangle-\langle{\vec k,r;\vec k',r'}|t_1+t_2|{\vec k,r;\vec
k',r'}\rangle$, where $t_i$ is the kinetic energy operator of
particle $i$. As argued by Moszkowski \cite{mosz} and Bethe \cite{bethe},
it is expected that the true ground-state wave
function of the nucleus containing correlations coincide
with the independent particle, or Hartree-Fock wave function, for interparticle
distances $r\geq r_{heal}$, where $r_{heal}\approx 1$ fm is the
so-called ``healing distance". This behavior is a consequence
of the constraints imposed by the Pauli Principle.
A natural consequence of having the
correlations introduced by an unitary operator is the occurrence of
a normalization constraint on $f(r)$,
\begin{equation}
\int~(f^2(r)-1)~d^3r=0.\label{c0}
\end{equation}

The correlated ground state energy becomes
\begin{eqnarray}
{\cal E}&=&\frac{\nu}{\pi^2}\int_0^{k_F} k^2 ~dk~
\left[|k|\sin\chi(k)~+~M\cos\chi(k)\right]
~+~\frac{\tilde F_\sigma(0)}{2}\rho_s^2 +\frac{\tilde
F_{\omega}(0)}{2}\rho_B^2\nonumber\\
&-&\frac{4}{(2\pi)^4} \int_0^{k_f} k^2~ dk~ {k'}^2~ d
k'~ \left\{\Big[ |k| \sin\chi(k)+ 2~M\cos\chi(k)\Big] I(k,k')
+ |k|~\sin\chi(k')~J(k,k') \right\}\nonumber\\
&+&\frac{1}{(2\pi)^4}\int_0^{k_f} k~ dk~ k'~ dk'
\left[\sum_{i=\sigma,\omega} A_i (k,k')
+\cos\chi(k)\cos\chi(k') \sum_{i=\sigma,\omega}B_i (k,k') 
+ \sin\chi(k)\sin\chi(k')
\sum_{i=\sigma,\omega} C_i(k,k')\right]\nonumber\\
\end{eqnarray}
where
$A_i$, $B_i$, $C_i$ , $I$ and $J$ are
exchange integrals. In the above equation for the energy density, the
first term results from the kinetic contribution, the second and third 
terms come respectively from the $\sigma$ and $\omega$ direct contributions
from the potential energy with correlations, the fourth  from the  exchange correlation contribution
from the kinetic energy, and the last one from the $\sigma +
\omega$ exchange contributions from the potential energy with
correlations. The direct correlation contribution is zero due to
(\ref{c0}). The
angular integrals are given by 
$A_i(k,k')=B_i(k,k')=2\pi ~g_i^2/4\pi\int_0^\pi d \cos \theta
~\tilde F_i(k,k',\cos \theta),$
$C_i(k,k')=2\pi ~g_i^2/4\pi\int_0^\pi \cos \theta ~d \cos \theta
~\tilde F_i(k,k',\cos \theta),$ 
$I(k,k')=2\pi \int_0^\pi d \cos \theta ~\tilde C_1(k,k',\cos \theta),$ and
$J(k,k')=2\pi \int_0^\pi \cos \theta ~d \cos \theta
~\tilde C_1(k,k',\cos \theta), $ 
where
\begin{equation}
\tilde F_i(\vec k,\vec k')=\int \left[f(r)V_\tau( r)f( r)\right]~
e^{i(\vec k-\vec k') \cdot \vec r}~d\vec r \quad \quad \mbox{and}\quad\quad
\tilde C_1(\vec k,\vec k')=\int (f^2(r)-1)~
e^{i(\vec k-\vec k') \cdot \vec r}~d\vec r.
\end{equation}
The baryon density and the scalar density are
\begin{equation}
\rho_B =\frac{2~\nu}{(2\pi)^3}
\int_0^{k_f} d{\vec k}=\frac{2~\nu ~k_f^3}{6\pi^2},\quad\quad
\rho_s = \frac{2~\nu}{(2\pi)^3}
\int_0^{k_f} \cos\chi(\vec k)~d{\vec k}.
\end{equation}
\begin{table}
\begin{ruledtabular}
\caption{Parameters and ground state properties of nuclear
matter at saturation density are given. These results were obtained with
fixed: $M=939$ MeV $m_\sigma=550$ MeV, $m_\omega=783$
MeV at $k_{F0}=1.3$ fm$^{-1}$ with binding energy
$E_B=\varepsilon/\rho-M=-15.75$ MeV. We have used a density dependent
parameter (HF+corr) $\gamma=600+400k_F/k_{F0}$ MeV for the correlation. 
} \label{tab}
\begin{tabular}{cccccccccccc}
&$g_\sigma$ & $g_\omega$&$\alpha$& $\beta$& $\gamma$& K &
$M^*/M$& ${\cal T}/\rho_B -M$& ${\cal V}_d/\rho_B$&
${\cal V}_e/\rho_B$&${\cal T}^C/\rho_B$ \\
& & & & (MeV) & (MeV)& (MeV)& & (MeV)&(MeV)&(MeV)&(MeV)\\
\hline
Hartree &11.079&13.806&  &  &  &  540 & 0.540 & 8.11 &-23.86 &  & \\
HF &10.432&12.223&  &  &  &  585 & 0.515 & 5.87 &-37.45 &15.83& \\
HF+corr&4.4559&2.6098& 13.855 &-2252.448 &1000 &429. &0.625&
15.95&-73.12 &20.46&19.96
\end{tabular}
\end{ruledtabular}
\end{table}
The couplings $g_\sigma$, $g_\omega$, the meson masses, $m_\sigma$
and $m_\omega$ and
also three more parameters from the short range correlation
function, $\alpha$, $\beta$ and $\gamma$ have to be fixed. The couplings 
are chosen so as to satisfy the ground state properties of
the nuclear matter. We choose 
$m_\sigma=550$ MeV and take $m_\omega=783$ MeV. 
The normalization condition (\ref{c0}) determines $\beta$. 
We fix $\alpha$ by minimizing
the energy. For the parameter
$\gamma$  we consider a function which increases linearly with the
Fermi momentum, of the form $\gamma=a_1+ a_2\, k_F/k_{F0}$ where $a_1$
and $a_2$ are free parameters. This is
consistent  with the idea that the healing distance
decreases as $k_F$ increases. 
\begin{center}
\begin{tabular}{ll}
\vspace {-2.8in}\\
\includegraphics[width=8.6cm]{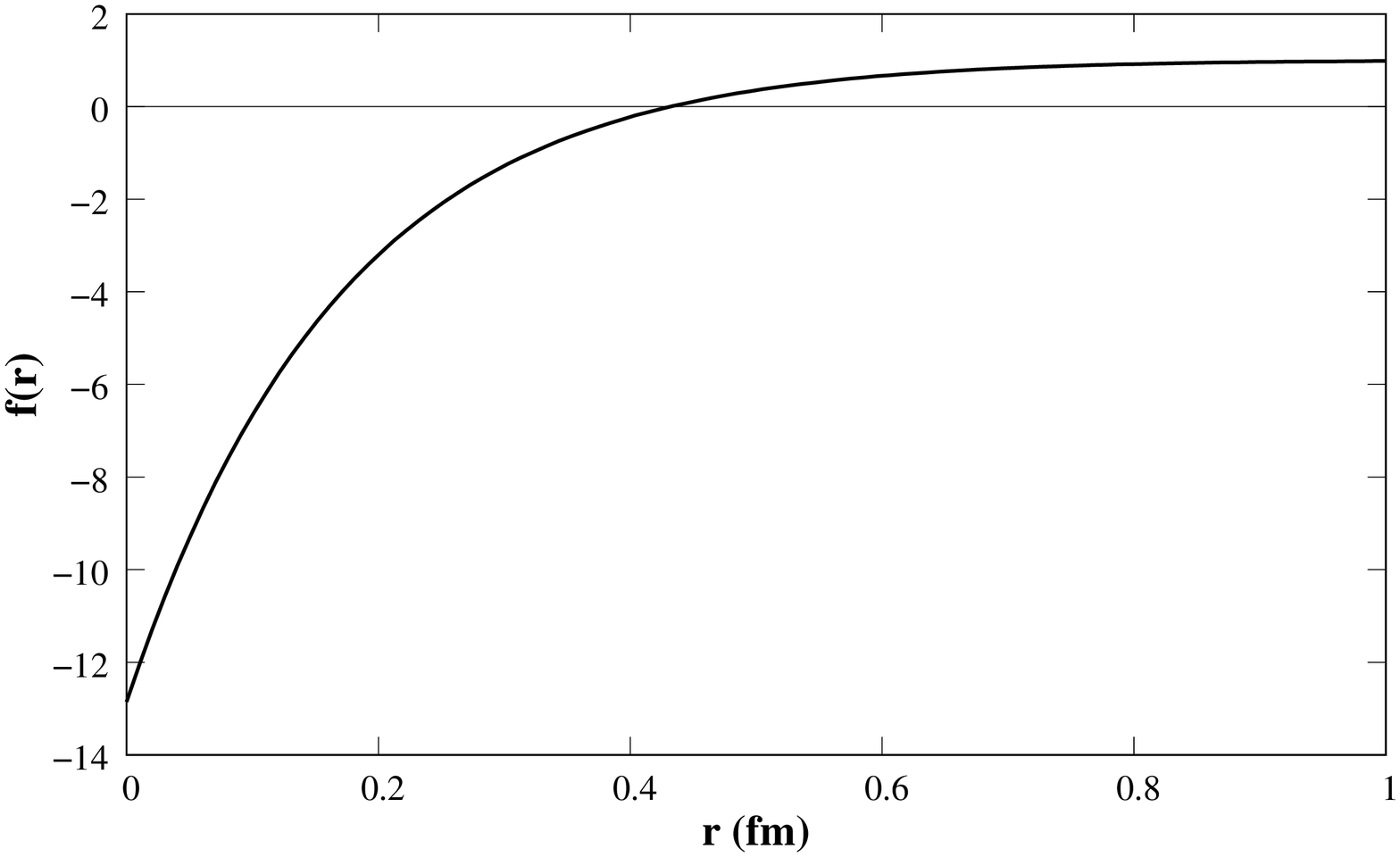}&
\includegraphics[width=9.cm]{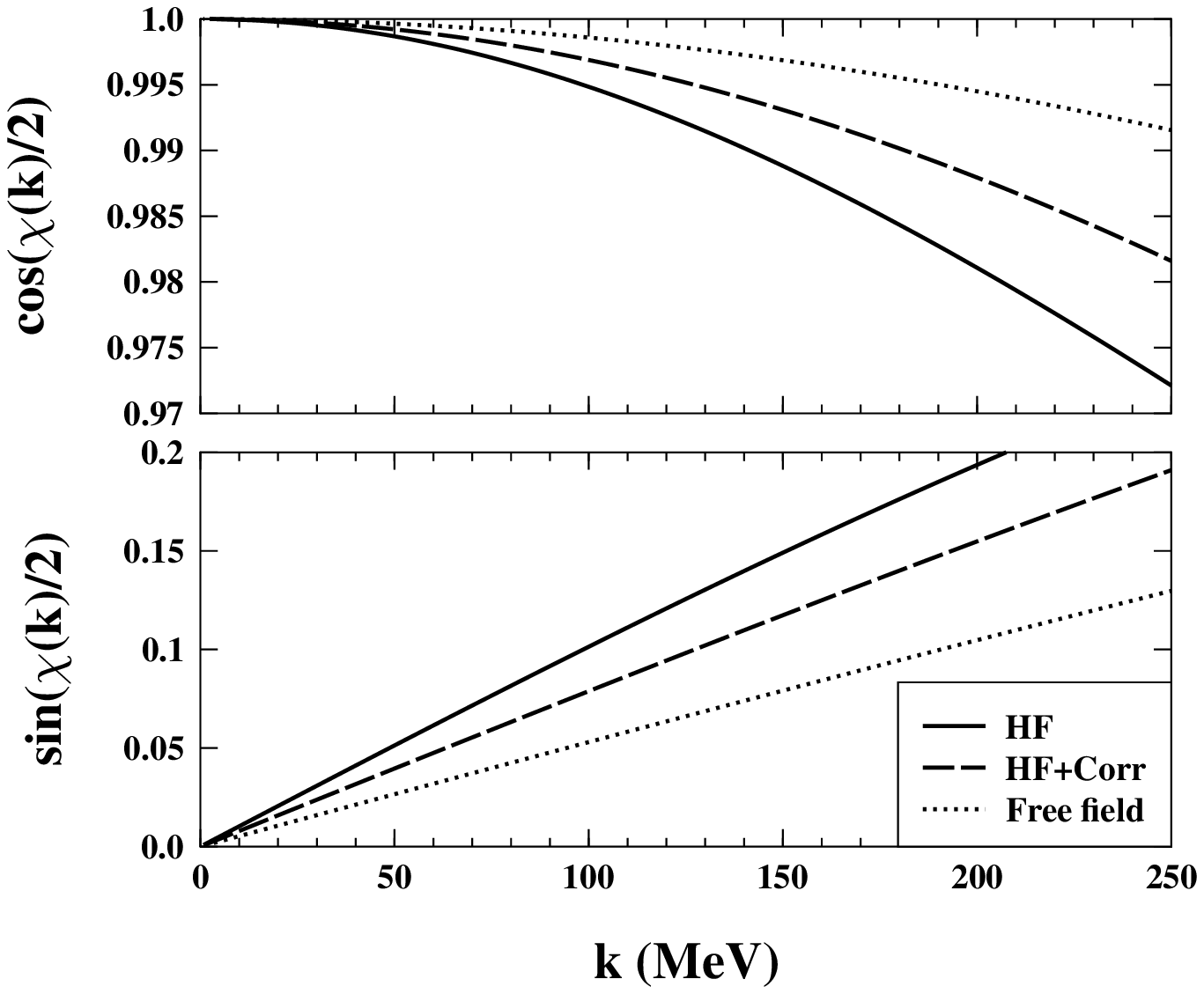}\\
~~~~~~~~~~~Fig. 1 The correlation function $f(r)$ &
~~~~~~~~~~~~~~~~~~~~~~Fig. 2 The variational angle $\chi(k)$ 
\end{tabular}
\end{center}
The correlation function $f(r)$ is plotted in figure 1 as a function
of the relative distance.
The inclusion of correlations introduces an extra node in the
ground-state wave-function contrary to what generally happens in
non-relativistic calculations with a hard core. In this case the wave
function has a wound. The quantities cos$(\chi(k)/2)$ and sin$(\chi(k)/2)$  
are plotted in figure 2. They show how the interaction and
correlations make the wave-function deviate from the free
wave-function, represented by a dotted line. The correlated angle $\chi(k)$
lies between the Hartree-Fock(HF) and the free wavefunction angles. As a 
consequence, we will see that the correlated effective mass will not 
decrease so fast with density as the HF effective mass.

In table \ref{tab}, we have tabulated the 
parameters used in our calculation together with the compressibility $K$,
the relative effective mass $M^*/M$, the kinetic energy ${\cal T}/\rho_B-M$,
the direct and exchange parts of the potential energy (${\cal V}_d/\rho_B$
and ${\cal V}_e/\rho_B$ respectively) with correlation and the correlation 
contribution to the 
kinetic energy ${\cal T}^C/\rho_B$, all calculated at the saturation point.
Notice that a HF calculation produces an EOS which is stiffer than the 
one obtained at the Hartree level. However, the inclusion of correlations gives a
larger effective mass than both Hartree and HF calculations and a softer EOS.  
In fact, the contribution of direct and 
exchange correlation terms are of the same order of magnitude of the other 
terms in the energy per particle. Hence, they cannot be disregarded.

We have computed the binding energies as function of the density for the 
Hartree, HF and HF+Corr and compared with the 
quark-meson-coupling model (QMC) \cite{qmc} and a non-linear Walecka model 
NL3 \cite{nl3}, as can be seen from fig. 3.
The inclusion of correlations make the equation of
state (EOS) softer than Hartree or HF calculations. NL3 and QMC
also provide softer EOS around nuclear matter saturation density
but around two times saturation density,  the EOS with
correlations is softer than NL3.
In figure 4 we plot the effective mass as a function of density.
If  correlations   are included the effective mass does not decrease
so fast with the increase of density as in a Hartree or,  even worse, HF
calculation. This explains the softer behavior of the EOS with correlations.

We conclude referring that, although correlation effects in the
Hartree and HF calculations are may be taken partially into account by a 
correct choice of
the coupling constants, the explicit introduction of correlations
has other effects such as softening the EOS.
\begin{center}
\begin{tabular}{ll}
\includegraphics[width=8.cm]{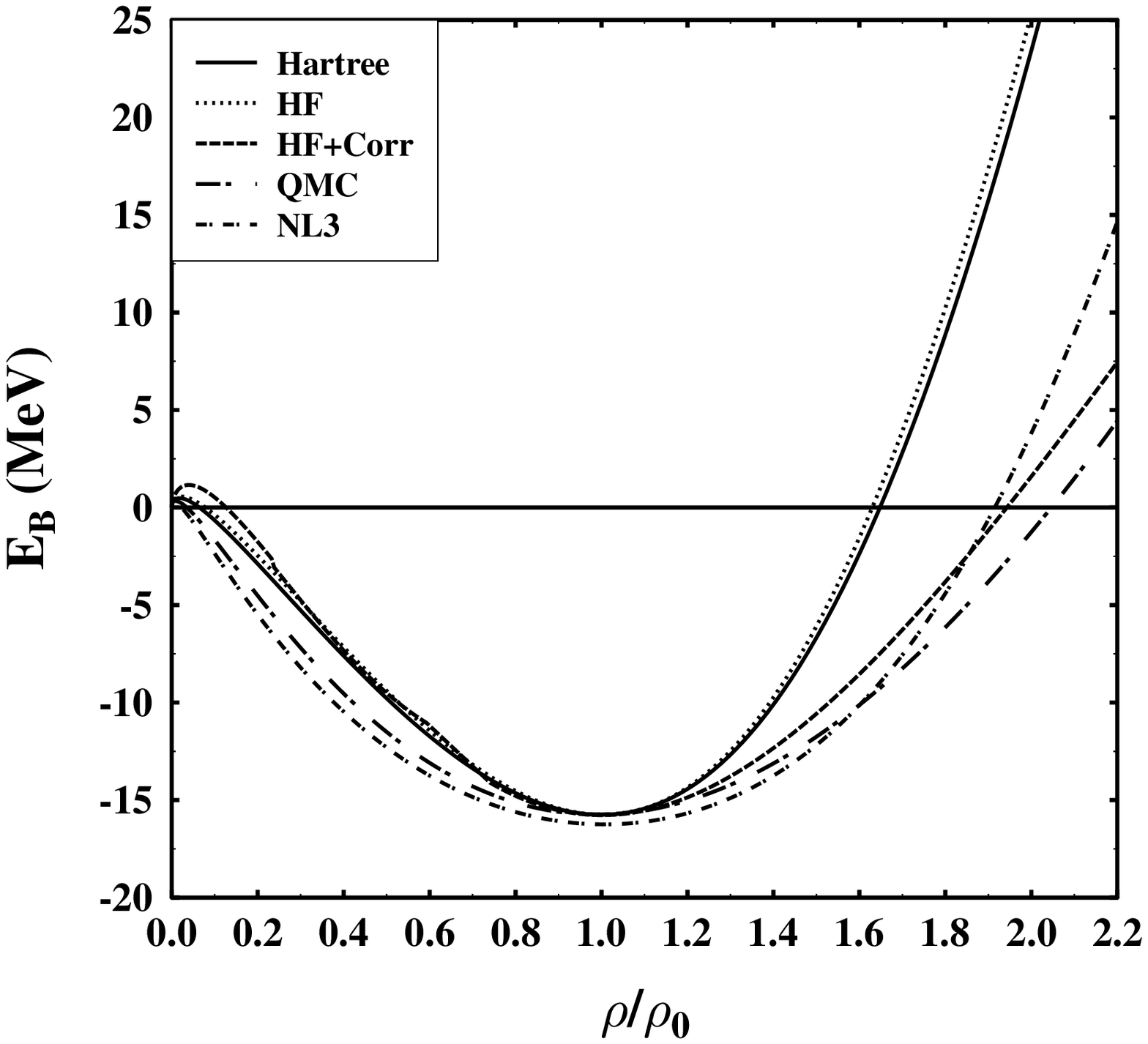}&
\includegraphics[width=8.cm]{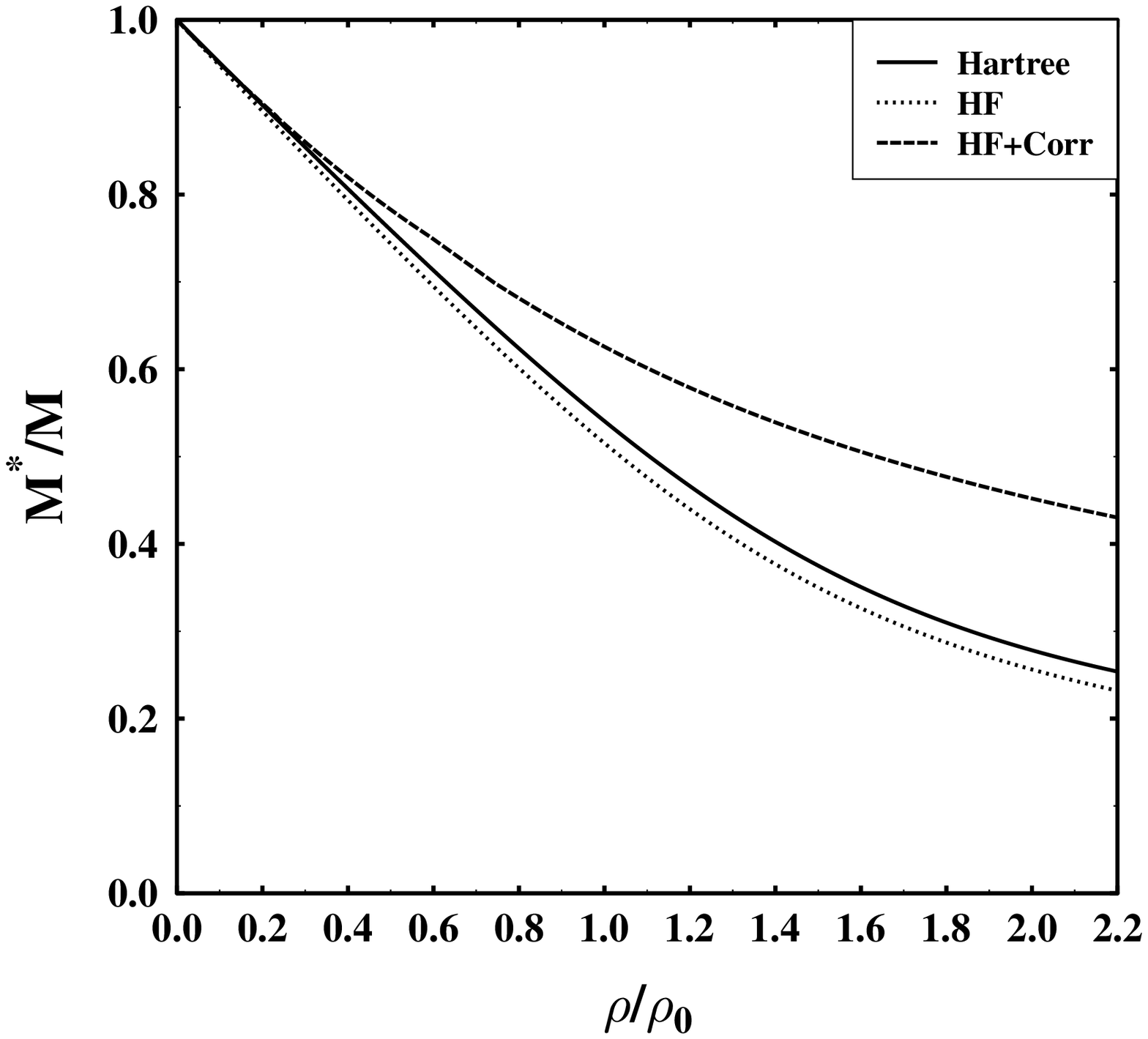}\\
\vspace {-1.0in}\\
~~~~~~Fig. 3 EoS for different parametrizations &
~~~~~~~~~~~~Fig. 4 Effective mass as a function of density 
\end{tabular}
\end{center}
\section*{ACKNOWLEDGMENTS}
This work was partially supported by CNPq (Brazil), CAPES (Brazil)/GRICES
(Portugal) under project 100/03,  FEDER and FCT (Portugal) under the
projects POCTI/FP/FNU/50326/2003, POCTI/FIS/451/94 and POCTI/FIS/35304/2000.

\end{document}